\documentclass[prd,showpacs,showkeys,nofootinbib,floatfix,eqsecnum,
               fleqn,preprint,12pt,tightenlines]{revtex4-1} %for preprint

\usepackage{xcolor,graphicx,hyperref}
\usepackage{amsmath,amssymb,revsymb,graphicx,dcolumn}

\newcommand{\beq}{\begin{equation}}
\newcommand{\eeq}{\end{equation}}
\newcommand{\beqa}{\begin{eqnarray}}
\newcommand{\eeqa}{\end{eqnarray}}
\newcommand{\bsubeqs}{\begin{subequations}}
\newcommand{\esubeqs}{\end{subequations}}

\newcommand{\sgn}{ \mathrm{sgn}}                % sign
\newcommand{\half}{{\textstyle \frac{1}{2}}}    % \frac12

\begin{document}

\begin{widetext}
%
%\noindent arXiv:2111.07962
%\hfill KA--TP--27--2021\;(\version)
%
\noindent Phys. Rev. D  \textbf{105}, 084066 (2022) \hfill    arXiv:2111.07962
%
%\noindent \hfill KA--TP--14--2020\;(\version)
%
\newline\vspace*{3mm}
\end{widetext}

\title{Big bang as a topological quantum phase transition}

\author{\vspace*{5mm} F.R. Klinkhamer}
\email{frans.klinkhamer@kit.edu}
\affiliation{Institute for Theoretical Physics,
Karlsruhe Institute of Technology (KIT),\\
76128 Karlsruhe,  Germany\\}

\author{G.E.~Volovik}
\email{grigori.volovik@aalto.fi}
\affiliation{Low Temperature Laboratory, Aalto University,  
P.O. Box 15100, FI-00076 Aalto, Finland \\
and\\
Landau Institute for Theoretical Physics, 
Academician Semyonov Avenue 1a,   
Chernogolovka 142432,  Russia\\}

%%\date{\today}

\begin{abstract}
\vspace*{1mm}\noindent
It has been argued that a particular type of quantum-vacuum variable $q$ 
can provide a solution to the main cosmological constant problem
and possibly also give a cold-dark-matter component.
We now show that the same $q$-field may suggest a new interpretation of
the big bang, namely as a quantum phase transition between topologically
inequivalent vacua. These two vacua are characterized by the
equilibrium values $q=\pm \, q_{0}$ and there is a kink-type solution
$q(t)$ interpolating between
$q=- q_{0}$ for $t\to -\infty$ and $q= +q_{0}$ for $t\to \infty$,
with conformal symmetry for $q=0$ at $t=0$.
\end{abstract}


%%mathematical and relativistic aspects of cosmology  98.80.Jk
%%Quantum phase transitions, 64.70.Tg, 05.30.Rt

\maketitle

\section{Introduction}
\label{sec:Introduction}

Several years ago, we have proposed a
condensed-matter-inspired approach
to the cosmological constant problem~\cite{Weinberg1988}.
Our approach goes under the name of
$q$-theory~\cite{KlinkhamerVolovik2008a,%   
KlinkhamerVolovik2008b,KlinkhamerVolovik2009,KlinkhamerVolovik2016-CCP,%
KlinkhamerVolovik2016-brane} and we present a brief review
in App.~\ref{app:Background-material}.
This ``$q$'' is a microscopic (high-energy) variable
of the quantum vacuum and
its macroscopic (low-energy) equations are Lorenz invariant
and governed by thermodynamics.
Later, we have also realized that rapid oscillations of the
$q$-field can act as a cold-dark-matter
component~\cite{KlinkhamerVolovik2016-q-DM,%
KlinkhamerVolovik2016-more-q-DM,KlinkhamerMistele2017,NittaYokokura2019}.

Now, we will discuss a third possible application of the
$q$-field, namely as an effective regulator of the big bang
singularity.
As $q$ is the variable that describes the deep quantum vacuum,
all coupling constants of the Standard Model,
as well as the gravitational coupling constant $G$, are functions of $q$.
An appropriate functional dependence
$G(q)$ may actually lead to a
kink-type behavior $q(t)$ of the vacuum variable and a
corresponding bounce-type behavior of the cosmic scale factor $a(t)$.
Our scenario replaces
the big bang singularity~\cite{HawkingEllis1973} of the Friedmann cosmology
by a topological quantum phase transition
(see, e.g., \mbox{Refs.~\cite{QPT-review1,QPT-review2,QPT-review3,QPT-review4}}
for four complementary reviews on the physics of
quantum phase transitions).

Topological matter~\cite{HasanKane2010,Wieder-etal2021}
and the topological quantum vacuum
are characterized by topological quantum numbers,
such as the Chern number, which is typically an integer.
For a continuous variation of the parameters of the system,
the topological vacuum can experience a
topological quantum phase transition
with a change in the value of the topological invariant.
As an integer invariant cannot vary continuously,
the intermediate state of a topological transition
may have special properties. For example,
if a discrete or continuous symmetry is broken
in the vacua on both sides of the transition, then
this symmetry can be restored in the intermediate state.
In particular, if the transition takes place between
two fully gapped (massive) vacua,   %%E
then the intermediate state
is gapless (massless)~\cite{QPT-review4}.
In our scenario, the ``big bang'' represents
such a specific intermediate state between two vacua
with nontrivial topology.
Here, the intermediate state is the trivial vacuum 
in which gravity is absent (i.e., $1/G=0$,
so that there is no Einstein--Hilbert term in the action)
and the conformal symmetry is restored.

In this reinterpretation of the ``big bang,'' the metric is kept
unchanged in the standard Robertson--Walker form
(unlike the different metric
used in Refs.~\cite{Klinkhamer2019,Klinkhamer2019-More,
KlinkhamerWang2019,KlinkhamerWang2020};
see Ref.~\cite{Klinkhamer2021-APPB-review} for a review
which also contains the original references of Friedmann
and others).
In fact, the result of the present paper improves upon
an earlier kink-bounce solution~\cite{Klinkhamer2020-Another},
which had a completely \emph{ad hoc} underlying theory.
Here, the underlying theory has a direct physical motivation,
as will become clear in the following.

%%\newpage%%tmp
\section{Anomaly-type term}
\label{sec:Anomaly-type-term}

The standard topological $\Theta$ term in the action
has the following form:
\begin{eqnarray}
S_\Theta = -\frac{1}{4\pi} \,
\epsilon^{\alpha\beta\gamma\delta}
\int d^4 x \, \Theta \, F_{\alpha\beta}F_{\gamma\delta} \,.
\label{eq:theta_conventional}
\end{eqnarray}
Here, there is a 2-form curvature $F=dA$
from a 1-form gauge field $A$.
Natural units with $\hbar=c=1$ are used throughout.
In topological vacua,
such as topological insulators~\cite{HasanKane2010},
$\Theta$ is determined
by a (quantized) topological invariant.
This topological invariant is, in fact, given by 
the second Chern number expressed
in terms of the Green's functions as an integral over the Brillouin zone
of the crystalline topological insulator;
see, e.g., Ref.~\cite{NissinenVolovik2019}.

The topological vacuum may also contain
higher-form gauge fields~\cite{Gaiotto-etal2015,Dubinkin2020}.
In the present paper, we focus on a 3-form gauge
field~\cite{3-form-A-first-batch,3-form-A-second-batch,3-form-A-third-batch},
for which the topological term reads~\cite{Gaiotto-etal2015}
\begin{eqnarray}
S_\phi = -\frac{1}{2\pi} \int \phi \,dA^{(3)}\,,
\label{eq:phi}
\end{eqnarray}
with a dimensionless (pseudo)scalar field $\phi$
[here, and in the following, we put ``pseudo'' in parentheses,
as we do not know the microscopic origin of the 3-form gauge field
and its transformation properties].
The main reason for considering
topological vacua with a 3-form gauge field is that,
as emphasized by Hawking~\cite{3-form-A-second-batch}, in particular,
the 3-form gauge field can perhaps solve the cosmological constant problem.

In our approach~\cite{KlinkhamerVolovik2008a,KlinkhamerVolovik2008b},
the parameter $\Theta$ or $\phi$ plays the role of
a dimensionless
chemical potential $\mu$ and (\ref{eq:phi}) becomes 
\begin{eqnarray}
S_\mu = -\frac{1}{24} \,
\epsilon^{\alpha\beta\gamma\delta}
\int d^4 x \, \mu \, F_{\alpha\beta\gamma\delta} \,.
\label{eq:theta4}
\end{eqnarray}
Using the proper normalization of the 3-form,
we can choose $\mu = \pm 1$.
In principle, we may consider $\Theta$ and $\mu$ as a dimensionless
(pseudo)scalar field $\chi$ with periodicity $\chi= \chi + 2\pi$.
But, here, we consider this
(pseudo)scalar $\mu$
as a constant, because it is the constant of motion of the dynamics.
Simultaneously, $\mu$ may serve as a topological invariant,
which changes abruptly  in the topological quantum phase transition
across the ``big bang.''

%%\newpage%%tmp
\section{Vacuum variable}
\label{sec:Vacuum-variable}

The vacuum variable $q$ can be written
in terms of the 4-form field strength
by the following expression:
\begin{eqnarray}
F_{\alpha\beta\gamma\delta} = q \, e \,
\epsilon_{\alpha\beta\gamma\delta}\,,
\label{eq:q}
\end{eqnarray}
where $e$ is the tetrad determinant and
$\epsilon_{\alpha\beta\gamma\delta}$ the completely antisymmetric
Levi--Civita symbol normalized to unity.
The (pseudo)scalar $q$ has a mass dimension of 4, 
while $\mu$ is dimensionless.
Remark that this vacuum variable $q$  differs from the one
used in our previous
papers~\cite{KlinkhamerVolovik2008b,KlinkhamerVolovik2016-CCP}
and App.~\ref{subapp:Action-and-field-equations-of-q-theory} here,
where the vacuum variable has a mass dimension of 2. 

The term \eqref{eq:theta4} now has the form
\begin{eqnarray}
S_\mu = - \int d^4 x \, e\, \mu \, q \,.
\label{eq:theta5}
\end{eqnarray}
Precisely this term $-\mu \,q$ enters
the Einstein gravitational equation and
cancels the vacuum energy density $\epsilon(q)$ in equilibrium for $q=q_{0}$, 
where $[d\epsilon/dq\,]_{q=q_{0}}=\mu$.
The thermodynamic (macroscopic) vacuum energy,
\beq
\label{eq:rhoV}
\rho_{V}(q) \equiv \epsilon(q) -\mu \,q\,,
\eeq
enters the Einstein equation as a cosmological-constant term
and self-adjusts to zero in equilibrium~\cite{KlinkhamerVolovik2008a}.
For the record, this Einstein equation is given
by \eqref{eq:EinsteinEquationF2}
in App.~\ref{subapp:Action-and-field-equations-of-q-theory},
where $q$ is denoted $F$.

Starting from the definition \eqref{eq:rhoV},
we recall the equilibrium
conditions~\cite{KlinkhamerVolovik2008a}:
\bsubeqs\label{eq:equilibrium-conditions}
\beqa
\label{eq:equilibrium-conditions-rhoV}
\rho_{V}(q_{0})&=&0 \,,
\\[2mm]
\label{eq:equilibrium-conditions-rhoVprime}
\rho_{V}^{\prime}(q_{0})&=&0\,,
\\[2mm]
\label{eq:equilibrium-conditions-rhoVprimeprime}
\rho_{V}^{\prime\prime}(q_{0})&>&0\,,
\eeqa
\esubeqs
where the prime stands for differentiation with respect to $q$.
Note
that \eqref{eq:equilibrium-conditions-rhoVprime}
fixes $\mu$ to the equilibrium value
\beq
\label{eq:mu0}
\mu_{0} \equiv
\left. \frac{\epsilon(q)}{d q}\right|_{q=q_{0}}\,.
\eeq
It remains an outstanding task
to find a proper microscopic
realization of $q$-theory that produces the
vacuum energy density $\epsilon(q)$
and the corresponding equilibrium values $q_{0}$ and $\mu_{0}$,
possibly with $\mu_{0}$ appearing as a (quantized) topological invariant.

%%\newpage%%tmp
\section{Quantum phase transition}
\label{sec:QPT}

\subsection{Setup}
\label{subsec:Setup}

If $\mu$ is a type of topological invariant, then $1/G$ should 
play the same role as the gap in the spectrum of a topological insulator:  
$\mu$ becomes zero at the transition between 
distinct topological vacua.
Conformal symmetry is restored at the transition point.
In this approach, the ``big bang'' would be represented as a
topological quantum phase transition from
the $\mu=+1$ equilibrium state to the $\mu=-1$ equilibrium state
via the trivial vacuum state with $q=0$ (and $\mu=0$).

In the trivial vacuum, both gravity and vacuum energy are absent,
and this vacuum obeys conformal symmetry.
In view of the mass dimension 4 of the $q$ variable and
the proper normalization of Newton's ``constant,''
the natural choice for the dependence of $1/G$ on $q$,
which is consistent with conformal symmetry, is as follows:
\begin{eqnarray}
\frac{1}{G(q)} = \sqrt{|q|}\,.
\label{eq:G}
\end{eqnarray}
It appears impossible to obtain an analytic
$q$ behavior of $1/G(q)$ without use of an energy scale $q_{0}>0$,
an example being $1/G(q)=q^2/(q_{0})^{3/2}$.
We prefer the \textit{Ansatz} \eqref{eq:G},
which is consistent with having conformal symmetry
at $q=0$.

For our cosmological discussion,
we take the standard spatially flat Robertson--Walker
metric~\cite{HawkingEllis1973},
\bsubeqs\label{eq:RW}
\beqa\label{eq:RW-ds2}
\hspace*{-0mm}
ds^{2}
&\equiv&
g_{\alpha\beta}(x)\, dx^{\alpha}\,dx^{\beta}
=
- d t^{2} + a^{2}(t)\;\delta_{ab}\,dx^{a}\,dx^{b}\,,
\\[2mm]
a(t) &\in& \mathbb{R}\,,
\\[2mm]
\hspace*{-0mm}
 t  &\in& (-\infty,\,\infty)\,,
 \quad
 x^{a} \in (-\infty,\,\infty)\,,
\eeqa
\esubeqs
where $t$ is the cosmic time coordinate given by $x^{0}=t$
and $a(t)$ the cosmic scale factor.
The spatial indices $a$, $b$ in \eqref{eq:RW-ds2}
run over $\{1,\, 2,\, 3 \}$.
The generalized Maxwell equation for the 4-form field strength
and the generalized Einstein equation then
give the following equations for $q(t)$ and
the Hubble parameter $H(t) \equiv [da(t)/dt]/a(t)$:
\bsubeqs
\label{eq:ODEs}
\begin{eqnarray}
\label{eq:ODEs-rhoVprime}
\frac{d\rho_{V}}{dq} &=& \frac{d(G^{-1})}{dq}
\left(\frac{d H}{dt} + 2\,H^{2}\right) \,,
\label{first}
\\[2mm]
\label{eq:ODEs-Friedmann}
\rho_{V} &=& G^{-1}\,H^2 + H\,\frac{d(G^{-1})}{dt}  \,,
\label{second}
\end{eqnarray}
\esubeqs
where the vacuum energy density $\rho_{V}(q)$ has been defined
in \eqref{eq:rhoV}.
The above equations appear as \eqref{eq:MaxwellFRW}
and \eqref{eq:EinsteinFRW-E00eq}
in App.~\ref{subapp:Action-and-field-equations-of-q-theory},
for $\rho_{M}=P_{M}=0$
[in these equations, $q$ is denoted by $F$
and the factor $8\pi/3$ there will be absorbed into $G(q)$ here].
It needs to be emphasized that the original expressions 
\eqref{eq:MaxwellSolution} and \eqref{eq:EinsteinEquationF2} 
in App.~\ref{subapp:Action-and-field-equations-of-q-theory} are

\emph{universal}~\cite{KlinkhamerVolovik2008a,KlinkhamerVolovik2008b},
that is, independent of the particular realization of the conserved variable $q$.

For a constant gravitational coupling,
$G(q)=(8\pi/3)\,G_{N}$, \eqref{eq:ODEs-rhoVprime} forces the
vacuum energy density to be a constant (assumed to be nonnegative),
$\rho_{V}(q) =\text{const} = \Lambda \geq 0$.
In addition, \eqref{eq:ODEs-Friedmann} reduces to the standard Friedmann equation, 
$H^2=(8\pi/3)\,G_{N}\,\Lambda$, with a cosmological constant
$\Lambda$ and no matter.
Equation \eqref{eq:ODEs-Friedmann} makes clear that
having a nonconstant gravitational coupling
$G(q)$ significantly modifies the structure of
the Friedmann equation. This observation contrasts
with what happens for the ``regularized'' metric of
Refs.~\cite{Klinkhamer2019,Klinkhamer2019-More},
which only changes the $H^2$ term of the
Friedmann equation by a multiplicative Jacobian term,
as shown by Eq.~(3.3a) of Ref.~\cite{Klinkhamer2019}.

%%\newpage%%tmp
\subsection{Vacuum energy}
\label{subsec:Vacuum-energy}

The general behaviour of a self-sustained vacuum
does not depend much
on the particular choice of the vacuum energy density
$\epsilon(q)$ in the action. There are only the following requirements:
$\epsilon(q)$ should be zero in the trivial vacuum ($q=0$)
and the trivial vacuum should be unstable towards the equilibrium vacuum
($q=q_{0} \ne 0$).
The simplest possible form is
\begin{eqnarray}
\frac{\epsilon(q)}{q_{0}} =-\frac{3}{2}\, \frac{q^2}{q_{0}^2}
+ \frac{1}{2} \,\frac{q^4}{q_{0}^4}   \,.
\label{eq:epsilon}
\end{eqnarray}
Here, $q_{0} >0$ is the magnitude of $q$
in equilibrium, so that the present Newton ``constant''
is given by $G_{N}=3/(8\pi)\,G(q_{0})$.
We have also used a particular
normalization, so that the equilibrium vacua
have chemical potentials $\mu = \pm 1$, which are
\emph{assumed}
to correspond
to topological quantum numbers of the vacuum.
The value of the chemical potential in the equilibrium vacuum
fully determines the coefficients $-3/2$ and $1/2$ in  (\ref{eq:epsilon}).

Note that the trivial vacuum with $q=1/G=0$ and the real vacuum
with $q=\pm q_{0}$
have the same thermodynamic vacuum energy:
$\rho_{V}(0)= \rho_{V}(q_{0})$.
This agrees with the multiple-point criticality  
principle~\cite{FroggattNielsen1996,Sidharth2016,Volovik2004}.
However, as distinct from the multiple-point criticality principle,  
one of these vacua is unstable.
Actually, we can also use another form
of $\epsilon(q)$, in which both vacua are locally stable.
In quantum liquids, our construction corresponds to
the coexistence of liquid and vapour at a nonzero pressure, 
while at zero pressure the gas phase is unstable towards the liquid phase.

\begin{figure}[t]
%\vspace*{-0mm}
\begin{center}
%%rename BB-as-TopQPT-2021-fig1-v2.eps  --> BB-as-TopQPT-fig1-v301.eps
%%       --> BB-as-TopQPT-fig1-v4.eps
\hspace*{0mm}    %%fig1-v1.eps=fig1-v2.eps
\includegraphics[width=1\textwidth]{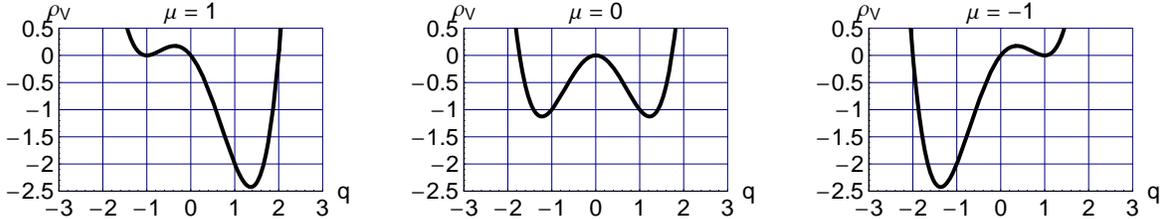}
%%{BB-as-TopQPT-fig1-v301.eps}
%%{BB-as-TopQPT-2021-fig1-v2.eps}
%%{BB-as-TopQPT-2021-fig1-v1.eps}
%%{BB-as-TopQPT-2021-fig1-v092.eps}
\end{center}\vspace*{-4mm}
\caption{Vacuum energy density  $\rho_{V}(q)$
from \eqref{eq:rhoV} for three different values
of the chemical potential, $\mu=1,\,0,\,-1$.
The $\epsilon(q)$ \textit{Ansatz} \eqref{eq:epsilon}
has been used and the scale $q_{0}$ has been set to unity.
According to \eqref{eq:equilibrium-conditions},
the stable equilibrium value of the vacuum variable is $q=-q_{0}=-1$ for
chemical potential $\mu=1$ and $q=q_{0}=1$ for $\mu=-1$.
For chemical potential $\mu=0$, the equilibrium state $q=0$ is unstable.
The left panel refers
to one phase of the quantum vacuum (characterized by $\mu=1$)
and the right panel
to another phase (characterized by $\mu=-1$).
}
\label{fig:rhoV}
\vspace*{0mm}
\end{figure}

Figure~\ref{fig:rhoV} shows,
for later reference, the vacuum energy density  $\rho_{V}(q)$
from \eqref{eq:rhoV} and \eqref{eq:epsilon}
at three different values of the chemical potential $\mu$
[the middle panel shows, in fact, the \textit{Ansatz}
function $\epsilon(q)$]

%%\newpage%%tmp
\subsection{Modified big bang}
\label{subsec:modBB}

Let us, first, find a solution at small $|t| \ll t_{0} \sim  q_{0}^{-1/4}$, 
i.e., for times shorter than the Planck time.
From \eqref{eq:ODEs} and \eqref{eq:epsilon},
we obtain for small positive $t$:
\bsubeqs
\begin{eqnarray}
 q&=&q_{0}\; \frac{t^2}{t_{0}^2} \,\,, \,\,  H = q_{0}^{1/2}\; \frac{t^2}{t_{0}} \,\,  , \,\,  G^{-1} = q_{0}^{1/2}\; \frac{t}{t_{0}}  \,,
\label{eq:22}
\\
\mu &=& -   1\,.
\label{eq:23}
\end{eqnarray}
\esubeqs
For later times, $t \gg t_{0}$, the solution will have
oscillations~\cite{KlinkhamerVolovik2008b}, which asymptotically 
approach the Minkowski vacuum with $|q|=q_{0}$.

There are now two possible extensions to negative $t$.
The first extension is symmetric with respect to time
reversal,
\bsubeqs
\label{eq:NegT11-NegT12}
\begin{eqnarray}
 q&=&q_{0}\; \frac{t^2}{t_{0}^2} \,\,, \,\,  H = q_{0}^{1/2}\; \frac{t^2}{t_{0}} \,\,  , \,\,  G^{-1} = q_{0}^{1/2}\; \frac{|t|}{t_{0}}  \,,
\label{eq:NegT11}
\\
\mu &=& -   1 \,, 
\label{eq:NegT12}
\end{eqnarray}
\esubeqs
and the second extension antisymmetric  
(see App.~\ref{app:Analytic-and-numeric-results} for further details),
\bsubeqs\label{eq:NegT21-NegT22}
\begin{eqnarray}
 q&=&q_{0}\; \frac{t\,|t|}{t_{0}^2} \,\,, \,\,  
 H = q_{0}^{1/2}\; \frac{t\,|t|}{t_{0}} \,\,  , \,\,  
 G^{-1} = q_{0}^{1/2}\; \frac{|t|}{t_{0}}  \,,
\label{eq:NegT21}
\\
\mu &=& -   \sgn \, t\,.
\label{eq:NegT22}
\end{eqnarray}
\esubeqs
The symmetric case \eqref{eq:NegT11-NegT12}
does not have topological stability, since the ``big bang'' connects 
two topologically equivalent vacua (each having $\mu =- 1$).
The antisymmetric case \eqref{eq:NegT21-NegT22}, on the other hand,
has the ``big bang''
(or, more precisely, the ``bounce'')
serving as a boundary between
different topological vacua (going, in Fig.~\ref{fig:rhoV},
from the left panel, via the middle panel, to the right panel),
and this boundary corresponds to the gapless boundary state of topological 
insulators.

In our case, the role of the gap
is played by the quantities $q$, $\mu$, $1/G$, $\epsilon$, and $\rho_{V}$, 
which are all
nullified in this boundary state.
Remark that this behavior of $1/G$, in particular,
is the opposite from that of
an earlier kink-bounce solution~\cite{Klinkhamer2020-Another},
which had  $G_\text{eff} \sim t^2$,
whereas here it is $1/G$ that vanishes at $t=0$, as well
as all energy densities.

%%\newpage%%tmp
\subsection{Possible physical interpretation}
\label{subsec:Possible-physical-interpretation}

We have presented further analytic and numeric results
in App.~\ref{app:Analytic-and-numeric-results}.
These results give the behavior 
for a cosmic time $t$ running over the whole real axis,
with an even solution $a(t)\geq 1$ having $a(t)\to\infty$
for $t\to \pm\infty$ and $a(0)=1$. 
The interpretation of this complete solution
is, however, rather subtle.
The crucial point is to realize that $t$
(or its dimensionless counterpart $\tau$) is merely
a time coordinate and not a ``physical'' time.
The ``thermodynamic''  time $\mathcal{T}$
(pressureless matter perturbations growing with increasing values
of $\mathcal{T}$) is, most likely, given by $|t|$. If we call
the  $t>0$ universe $U$ and
the  $t<0$ universe $\overline{U}$,
then the solution of App.~\ref{app:Analytic-and-numeric-results} 
corresponds to two universes (or possibly a universe-antiuniverse pair)
with local thermodynamic times:
\bsubeqs\label{eq:mathcalTU-mathcalTUbar}
\beqa
\label{eq:mathcalTU}
\mathcal{T}_{\,U} &=& t\,,  \phantom{-} \;\;\text{for}\;\;t>0\,,
\\[2mm]
\label{eq:mathcalTUbar}
\mathcal{T}_{\,\overline{U}} &=& -t\,, \;\;\text{for}\;\;t<0\,.
\eeqa
\esubeqs
A brief discussion of the ``thermodynamic''  time
is given in the last paragraph of Sec.~IV
of Ref.~\cite{KlinkhamerWang2020}
and a brief discussion of the two-universes
interpretation in the second point of Sec.~3.2
of Ref.~\cite{Klinkhamer2021-APPB-review}.
The scenario of a universe-antiuniverse pair
was first considered in Ref.~\cite{BoyleFinnTurok2018},
with a follow-up paper in Ref.~\cite{BoyleTurok2021a}.
The strict validity of the
papers~\cite{BoyleFinnTurok2018,BoyleTurok2021a}
is, however, rather doubtful,
as the Einstein gravitational equation used may not hold for
an infinite matter energy density
and an infinite Kretschmann curvature scalar~\cite{Klinkhamer2019}.

So, the physical picture we have is as follows.
The ``big bang'' is replaced by a topological quantum phase transition
\emph{in the vacuum}, which corresponds to a
``temporal'' kink-type solution of the vacuum variable $q(t)$.
(Recall that the standard kink solution
has the scalar field changing in a spacelike direction.)
The center of the kink-type solution requires the
strict absence of matter.
Matter may be generated by the oscillations
of the vacuum variable $q$ 
as discussed in App.~\ref{app:Analytic-and-numeric-results}. 
It may, however, be that the proper physical  
interpretation of the solution obtained 
in App.~\ref{app:Analytic-and-numeric-results} is not that of 
a bouncing cosmology (starting from
$t=-\infty$, passing through a bounce at $t=0$,
and proceeding on to $t=\infty$), but rather that
of a creation process at $t=0$, with two
more or less equivalent branches:
$U$ for $t>0$ and $\overline{U}$ for $t<0$.

%%\newpage%%tmp
\section{Outlook}
\label{sec:Outlook}

The question arises as to the origin of the 3-form gauge field,
which appears to be
essential for a proper description of the quantum vacuum.
One possible answer to this question may come from
the discussion in Ref.~\cite{BoyleTurok2021b}.
It is shown there that, in order to cancel all quantum anomalies,
we may need, in addition to the fundamental fermionic and gauge fields,
also certain scalars with mass dimension zero.
Our chemical potential $\mu$ is conjugate to  
the 4-form field strength $F$
(as discussed in Secs.~\ref{sec:Anomaly-type-term}
and \ref{sec:Vacuum-variable}) and
may be an example of such a dimensionless field.
Recall that the (pseudo)scalar $q$ follows from
the 4-form field strength $F$ by the expression \eqref{eq:q}.

On the other hand, if the ``big bang'' is regular,
i.e., fully determined by the dynamics of $q$ and the metric,
then the chemical potential is a dynamical invariant of these equations
and thus cannot change sign in this regular ``big bang.''
Then it would make sense
to consider the ``big bang'' with an antisymmetric tetrad determinant, $e(t)=-e(-t)$,
as discussed in
Refs.~\cite{BoyleFinnTurok2018,BoyleTurok2021a,Volovik2019}.
In this case, $q(t)$ can change sign with a fixed value of the chemical potential.
However, this scenario would require
an extra  equation for $e(t)$. For that reason,
it would seem preferable to use a theory in which $e(t)$ is the order parameter
of the symmetry breaking phase transition, so that
$e(t)$ would play the role of the vacuum variable
(cf. Refs.~\cite{KlinkhamerVolovik2019,Zubkov2019,%
KlinkhamerVolovik2021-in-preparation}).

In the discussion of the previous paragraph,
we have implicitly considered the
evolution from $t=-\infty$ to $t=+\infty$ (or \textit{vice versa}),
but it is also possible that a universe-antiuniverse pair
``starts'' at $t=0$, as discussed
in Sec.~\ref{subsec:Possible-physical-interpretation}
and perhaps in line with remark (ii)
in the last section of Ref.~\cite{BoyleTurok2021a}.
There is then a creation process at the coordinate value $t=0$,
which gives rise to a universe $U$ with chemical potential $\mu=-1$
(for an appropriate normalization)
and an antiuniverse $\overline{U}$ with chemical potential $\mu=1$.
In fact, there would be an emergent spacetime at $t=0$
with the proper matter fields and constants relevant to both branches
$U$ and $\overline{U}$, which each
would have a thermodynamic time $\mathcal{T}$
as given by \eqref{eq:mathcalTU}
and \eqref{eq:mathcalTUbar}, respectively.

One possible origin for such an emergent spacetime
(both its points and its metric) would be the IIB matrix
model~\cite{IKKT-1997,Aoki-etal-review-1999,Klinkhamer2021-master},
with further references in
the review~\cite{Klinkhamer2021-APPB-review}.
The interest of the IIB matrix model is that it appears
to provide an ``existence proof'' for the idea of
emergent spacetime and matter, but what new physics
really replaces Friedmann's big bang singularity
remains an open question.
Perhaps the main result of the present paper
provides a clue, namely that we have obtained
a kink-type solution $q(t)$ of 
the vacuum variable with a vanishing Hubble parameter $H$  
precisely at the central point of the kink-type solution
with $q=0$ and conformal symmetry.

\begin{acknowledgments}
We thank the referee for constructive remarks.
The work of G.E.V. has been supported by 
the European Research Council (ERC)
under the European Union's Horizon 2020 
research and innovation programme (Grant Agreement No. 694248).

\end{acknowledgments}

\begin{appendix}

%%\newpage%%tmp
\section{Background material}
\label{app:Background-material}

In this appendix, we provide some background
on a particular condensed-matter-physics approach ($q$-theory)
to the cosmological constant problem (CCP).

\subsection{CCP from the condensed-matter point of view}
\label{subapp:CCP-from-the-condensed-matter-view}

In condensed matter physics, we know both the infrared (IR) low-energy limit 
described by the effective quantum fields and the ultraviolet (UV)
high-energy limit corresponding to the atomic physics,
the analog of Planck-scale quantum gravity. That is why we can explicitly 
see how the zero-point energy of the effective quantum fields in the IR 
are completely canceled 
by the UV degrees of freedom in the ground state of any condensed matter system. 
This occurs due
the general thermodynamic Gibbs--Duhem relation,
which is applicable to any system, be it the relativistic quantum vacuum of 
elementary particle physics or the nonrelativistic equilibrium states of 
condensed matter physics. Such a natural cancellation of the vacuum energy 
in the relativistic quantum vacuum 
is demonstrated by our $q$-theory formalism~\cite{KlinkhamerVolovik2008a}.

The interplay of IR and UV physics in condensed matter systems can be 
clarified by the example of quantum liquids, such as liquid $^3$He and 
liquid $^4$He~\cite{Volovik2009}.  The ground state of each liquid may 
serve as a nonrelativistic analog of the relativistic quantum vacuum. 
The stability of the many-body system is supported by the conservation law 
for the atoms of the liquid. The energy of the many-body system 
is proportional to the number $N$ of atoms and to the volume $V$. 
These quantum liquids are
self-sustained systems. This means that, as distinct from gases where  
equilibrium states may exist only under external positive pressure, 
liquids have a nonzero equilibrium density even in the absence of an environment.

At the UV scale, the quantum liquid is the many-body state of $N$ atoms,
which are described by the nonrelativistic many-body Hamiltonian
(for simplicity, we consider liquid $^4$He with spinless atoms):
\begin{equation}
\mathcal{H}= -\frac{\hbar^2}{2m}\sum_{i=1}^N {\partial^2\over \partial{\bf r}_i^2}
+\sum_{i=1}^N\sum_{j=i+1}^N U({\bf r}_i-{\bf r}_j)\,.
\label{eq:TheoryOfEverythingOrdinary}
\end{equation}
Here, $m$ is the bare mass of the $^4$He atom
and $U({\bf r}_i-{\bf r}_j)$ the pair-interaction potential
of the bare atoms $i$ and $j$. The Hamiltonian
\eqref{eq:TheoryOfEverythingOrdinary} acts on the many-body wave  function  
$\Psi({\bf r}_1,{\bf r}_2, \ldots  , {\bf r}_i,\ldots  ,{\bf r}_j,\ldots )$.

In the thermodynamic limit $N\rightarrow \infty$, 
the many-body physics can be described in the second-quantized form, 
where the above Schr\"{o}dinger many-body Hamiltonian becomes the 
Hamiltonian of a quantum field theory:
\begin{eqnarray}
\widetilde{\mathcal{H}}&=&\mathcal{H}-\mu\, \mathcal{N}
\nonumber
\\
&=&
\int d{\bf x}\,\psi^\dagger({\bf x})\left[-{\nabla^2\over 2m}
-\mu \right]\psi({\bf x})
 +\frac{1}{2}\int d{\bf x}d{\bf y}\,U({\bf x}
 -{\bf y})\psi^\dagger({\bf x}) \psi^\dagger({\bf y})\psi({\bf y})\psi({\bf x})\,.
\label{eq:TheoryOfEverything}
\end{eqnarray}
For the $^4$He liquid,
$\psi({\bf x})$ is a bosonic quantum field, the  annihilation operator 
of the $^4$He atoms. Note an important difference between the atomic 
many-body Hamiltonian $\mathcal{H}$ in (\ref{eq:TheoryOfEverythingOrdinary}) 
and the quantum field theory Hamiltonian $\widetilde{\mathcal{H}}$ 
in (\ref{eq:TheoryOfEverything}).
The latter
contains, namely, a term with the chemical potential $\mu$,
which is the Lagrange multiplier
responsible for the conservation of particle number
$N=\int d^3r \,\psi^\dagger \psi$.

In the thermodynamics of liquids, the energy density is a function 
of the number density of atoms $n=N/V$, i.e.,  $<\mathcal{H}>= \epsilon(n)\, V$, 
while $\mu$ is the chemical potential which is thermodynamically conjugate to $n$, 
i.e., $\mu=d\epsilon/dn$. The vacuum energy in the quantum field theory of liquids, 
as described by (\ref{eq:TheoryOfEverything}), is
$<\widetilde{\mathcal{H}}>= \rho_{V}(n)\, V$, 
where the relevant vacuum energy density is $\rho_{V}(n)=\epsilon(n) - \mu\, n$.
From the general thermodynamic Gibbs--Duhem relation
at zero temperature, $\epsilon(n) = \mu\, n - P$, follows that,
in the ground state of the liquid, the energy density $\rho_{V}(n)$ has  
the following  equation of state:
\begin{equation}
\rho_{V}(n) \equiv \epsilon(n) - \mu\, n= -P\,.
\label{eq:EquationOfState}
\end{equation}
This equation of state does not depend on the microscopic structure 
of the liquid and on the detailed form of the function $\epsilon(n)$,
except for the required stability condition $d^2 \epsilon/dn^2>0$ 
in the equilibrium state. This holds for any many-body system 
in the limit $N\rightarrow \infty$. Hence, it is not surprising 
that the same equation of state $\rho_{V}=-P$ is applicable to 
the energy density of the relativistic quantum vacuum, which is the reason
why we have used the suffix $V$ on the energy density $\rho_{V}(n)$.

One important property of the liquid is that it is a self-sustained system: 
liquids are stable at zero external pressure, whereas gasses are not.
That is why, in the absence of an external environment, and thus at $P=0$, 
the relevant energy density in the ground state of the liquid is exactly zero, 
$\rho_{V}=0$. The contribution of $\epsilon(n)$ to $\rho_{V}(n)$ is
precisely  canceled by the contribution $-\mu\, n$ without any fine-tuning. 
This is a direct consequence of the laws of thermodynamics.
The equilibrium values of $n$ and $\mu$ in the ground state (vacuum) 
of the liquid are determined by
the following equations:
\begin{equation}
\epsilon(n_{0}) - \mu_{0}\, n_{0}=0 \,,
 \quad
 \mu_{0} = \left.\frac{d\epsilon}{dn}\right|_{n=n_{0}}\,.
\label{eq:Equilibrium}
\end{equation}

Let us turn to the interplay of the IR and the UV. The quantity $\epsilon(n)$ 
is determined by atomic physics, i.e., the physics of the UV. On the other hand, 
the quantity $\rho_{V}(n)$ belongs to the IR physics,
where two UV contributions cancel each other in full equilibrium.
This means that, for deviations from equilibrium or at nonzero temperature, 
the energy density $\rho_{V}$ is determined by effective theories in the IR. 
The IR physics contains, in particular, bosonic or fermionic quasiparticles, 
which represent the analog of matter on the
quantum-vacuum background of the liquid. If the temperature is nonzero, 
then these quasiparticles contribute to the equation of state, making for
$\rho_{V} \neq 0$.
For the case of a linear spectrum of these quasiparticles,
the vacuum energy density is comparable with the free energy 
of the quasiparticles, $\rho_{V} \propto T^4$.
This situation resembles the one of our present Universe,
where the numerical value of the vacuum energy density is of the same order
of magnitude as that of the energy density of matter.

Now about divergences in the effective field theory of the quantum liquids. 
In the low-energy limit, the quantum liquids contain fermionic and bosonic 
quasiparticles in the background of the quantum vacuum. They play the role of 
matter and are described by effective quantum fields. These fields have the 
conventional zero-point energies, which give rise to (negative or positive) 
divergent contributions to the vacuum energy.
In elementary particle physics, these divergences require
consideration of the UV physics.
But, in quantum liquids,
we know the UV (atomic) physics with its Hamiltonians 
 (\ref{eq:TheoryOfEverythingOrdinary}) and (\ref{eq:TheoryOfEverything}), 
 and we also know that the nullification of the vacuum energy
in full equilibrium is protected by thermodynamics. 
That is why such divergences are natural but not catastrophic: 
the UV degrees of freedom also obey the laws of
thermodynamics, and the divergent terms coming from the emergent fields are 
naturally canceled in the equilibrium vacuum by the atomic (trans-Planckian) 
degrees of freedom.

The above discussion demonstrates that thermodynamics
is more general than relativistic invariance, which suggests that the 
cosmological constant problem can be studied by
using our experience from condensed matter physics.

%%\newpage%%tmp
\subsection{CCP in $q$-theory}
\label{subapp:CCP-in-q-theory}

Let us, then, apply condensed-matter insights
to the relativistic quantum vacuum.
Actually, the only requirement of the relativistic vacuum is that it should belong to the class of self-sustained media. This is what distinguishes the condensed-matter approach from other possible scenarios for the nullification of the cosmological constant in full equilibrium.
The vacua of this class of self-sustained systems
can be characterized by a particular vacuum variable, the conserved quantity which we have denoted by $q$.
This $q$ is similar to the density of atoms $n$,
but $q$ does not violate the relativistic invariance of the vacuum~\cite{KlinkhamerVolovik2008a}. The vacuum equation of state in terms of $q$ has the same form as (\ref{eq:EquationOfState}):
\begin{equation}
\rho_{V}(q) = \epsilon(q) - \mu\, q= -P\,,
\label{eq:EquationOfStateQ}
\end{equation}
where  $\mu$ is the chemical potential
corresponding to the conservation law obeyed by the vacuum ``charge'' $q$.

The expression (\ref{eq:EquationOfStateQ}) is general, as
it does not depend on the exact form of energy density $\epsilon(q)$, which is determined by the UV physics, Planckian or trans-Planckian. The field equation for the $q$-field and the Einstein equation for gravity interacting with the $q$-field have the general form as given
in App.~\ref{subapp:Action-and-field-equations-of-q-theory}
(where $q$ is denoted $F$).
It is important that the role of the cosmological ``constant''
in the Einstein equation is played by the IR quantity $\rho_{V}(q)$,
rather than by the UV energy density $\epsilon(q)$,
see \eqref{eq:EinsteinEquationF2} and \eqref{eq:EinsteinFRW-rhoV} below,
with $q$ denoted by $F$.
This demonstrates that the Einstein equation belongs to the 
class of  effective theories emerging in the low-energy corner.

Just as other members of the class of   
self-sustained vacua, the relativistic quantum vacuum 
may exist without external environment, i.e., 
at zero external pressure. The equilibrium values 
of $q$ and $\mu$ in the self-sustained relativistic 
vacuum are given by expressions
similar to those in (\ref{eq:Equilibrium}):
\begin{equation}
 \epsilon(q_{0}) - \mu_{0}\, q_{0}=0 \,,
\quad
\mu_{0} = \left. \frac{d\epsilon}{dq}\right|_{q=q_{0}}\,.
\label{eq:EquilibriumRelativistic}
\end{equation}

As these equations
do not depend on the underlying microscopic theory,
we can exploit different choices for the vacuum variable.
For example, we can choose as the vacuum variable the scalar $q$
from the 4-form field
strength~\cite{3-form-A-first-batch,3-form-A-second-batch,3-form-A-third-batch}.  This 4-form field was used for the CCP by
Hawking~\cite{3-form-A-second-batch}, in particular.
In Ref.~\cite{KlinkhamerVolovik2008a},
we extended the Hawking approach beyond the quadratic term, which allowed us to consider self-sustained quantum vacua.  It is very well possible
that the 4-form field not only serves as a toy model for the vacuum field
but that it represents a genuine fundamental field of
the quantum vacuum.

At this point, let us stress the main difference between
the scalar $q$ from the 4-form field strength
[see \eqref{eq:Fdefinition2} below, with $q$ denoted by $F$]
and a fundamental scalar field $\phi$.
While the fundamental scalar field $\phi$, coupled to the metric,
may produce a solution to the CCP, Weinberg’s ``no-go theorem''~\cite{Weinberg1988}
shows that the full nullification of the vacuum energy in equilibrium needs to be fine-tuned.
The scalar $q$ from the 4-form field strength
does not require fine-tuning. As distinct from the fundamental scalar field $\phi$,
the scalar $q$ from the 4-form field strength has the analog of the chemical potential $\mu$, which is responsible for the cancellation of the UV terms in (\ref{eq:EquilibriumRelativistic}).
Further details on the different roles of $q$ and $\phi$
are given in Sec.~2 of Ref.~\cite{KlinkhamerVolovik2016-brane}.

Let us mention, that the $q$-field approach
differs from unimodular gravity, which also generates a constant of
motion~\cite{Henneaux1989}. In the realization of $q$ theory in terms 
of the 4-form field strength, 
the theory is fully diffeomorphism invariant. For this thermodynamic approach,  
the constant of motion is the chemical potential $\mu$  corresponding to the 
conserved quantity $q$. In the equilibrium state of the system, both the 
chemical potential $\mu$ and the temperature $T$ are constant.

In principle, the tetrad determinant can also serve as a type 
of vacuum variable~\cite{KlinkhamerVolovik2019,Zubkov2019,%
KlinkhamerVolovik2021-in-preparation}.
This would correspond to the nonlinear extension of unimodular gravity, 
where the tetrad determinant (and, thus, the metric determinant) 
is not fixed, and all physical quantities are functions of this variable.

%%\newpage%%tmp
\subsection{Action and field equations of $q$-theory}
\label{subapp:Action-and-field-equations-of-q-theory}

%%based on vacvar-equilibration_v6.tex = PRD-78-063528(2008)

Here, we recall the main equations from
Ref.~\cite{KlinkhamerVolovik2008b}, where the
$q$-type vacuum variable is denoted $F$.
Indeed, the use of the notation $F$ in this appendix
makes clear that a particular
realization of $q$ is being considered, namely a
realization based on the 4-form field
strength $F_{\alpha\beta\gamma\delta}$
from a 3-form gauge field $A_{\beta\gamma\delta}$
(details will be given shortly).
The mass dimension of this particular vacuum variable $F$
is 2, whereas the  vacuum variable $q$ of
Secs.~\ref{sec:Vacuum-variable} and \ref{sec:QPT}
has a mass dimension of 4.

The starting point is the action as discussed in
our original paper~\cite{KlinkhamerVolovik2008a}, but
with Newton's constant $G_{N}$ replaced by
a gravitational coupling parameter $G$ which is taken to
depend on the state of the vacuum and thus on the vacuum variable $F$.
The $G(F)$ dependence is natural and must, in principle,
occur in the quantum vacuum.
Note that this $G(F)$ dependence allows
the cosmological ``constant'' to change with time, which is otherwise
prohibited by the Bianchi identities
and energy-momentum conservation (this point will be
discussed further in the last paragraph of the present appendix).

For natural units with $\hbar=c=1$, the action
considered takes the following form~\cite{KlinkhamerVolovik2008b}:
\bsubeqs\label{eq:EinsteinF-all}
\beqa
S[g,\, A,\, \psi]&=& -
\int_{\mathbb{R}^4} \,d^4x\, \sqrt{-g}\,\left(\frac{R}{16\pi G(F)} +
\epsilon(F)+\mathcal{L}^{M}(\psi)\right)\,,
\label{eq:actionF}
\\[2mm]
F_{\alpha\beta\gamma\delta} &\equiv& \nabla_{[\alpha}A_{\beta\gamma\delta]}\,,
\quad
F^2 \equiv - \frac{1}{24}\, F_{\alpha\beta\gamma\delta}\,F^{\alpha\beta\gamma\delta}\,,
\label{eq:Fdefinition}
\\[2mm]
F_{\alpha\beta\gamma\delta}&=&
F\,\epsilon_{\alpha\beta\gamma\delta}\,\sqrt{-g}\,,
\quad
F^{\alpha\beta\gamma\delta}=
F \,\epsilon^{\alpha\beta\gamma\delta}/\sqrt{-g}\,,
\label{eq:Fdefinition2}
\eeqa
\esubeqs
where a square bracket around spacetime indices denotes
complete antisymmetrization and $\nabla_{\alpha}$ is the covariant derivative.
The right-hand side of \eqref{eq:actionF}
shows only the functional dependencies on $F=F[A,g]$ and $\psi$,
while keeping the functional dependence on the metric $g$ implicit.
The field $\psi$ in \eqref{eq:actionF} stands
for a generic low-energy matter field
with a Lagrange density $\mathcal{L}^{M}$,
which is assumed to be without direct $F$--field dependence  
(it is possible to relax this assumption by changing the low-energy
constants in $\mathcal{L}^{M}$  to $F$--dependent parameters).  
The metric signature is taken as $(-+++)$.

Using \eqref{eq:Fdefinition2},
the variation of the action \eqref{eq:actionF} over the 3-form
gauge field $A$ gives the generalized Maxwell equation
in the following form~\cite{KlinkhamerVolovik2008b}:
\begin{equation}
\partial_{\alpha} \left( \frac{d\epsilon(F)}{d F}+\frac{R}{16\pi}
\frac{dG^{-1}(F)}{d F} \right) =0\,.
\label{eq:Maxwell2}
\end{equation}
The solution is simply
\begin{equation}
 \frac{d\epsilon(F)}{d F}+\frac{R}{16\pi} \frac{dG^{-1}(F)}{d F} =\mu \,,
\label{eq:MaxwellSolution}
\end{equation}
with an integration constant $\mu$.
The constant $\mu$ can be interpreted as a
chemical potential corresponding to the conservation law obeyed by
the vacuum ``charge'' $q\equiv F$; for further discussion,
see Refs.~\cite{KlinkhamerVolovik2008a,KlinkhamerVolovik2008b}
and Apps.~\ref{subapp:CCP-from-the-condensed-matter-view}
and \ref{subapp:CCP-in-q-theory} here.

Using \eqref{eq:MaxwellSolution},
the variation of the action \eqref{eq:actionF}
over the metric $g_{\alpha\beta}$ gives
the generalized Einstein equation
in the following form~\cite{KlinkhamerVolovik2008b}:
\begin{equation}
\frac{1}{8\pi G(F)}\Big( R_{\alpha\beta}-\half\,R\,g_{\alpha\beta} \Big)
+ \frac{1}{8\pi}
\Big( \nabla_{\alpha}\nabla_{\beta}\, G^{-1}(F)
- g_{\alpha\beta}\, \Box\, G^{-1}(F)\Big)
-\Big(\epsilon(F)-\mu\, F \Big)\, g_{\alpha\beta}
+T^{M}_{\alpha\beta} =0\,,
\label{eq:EinsteinEquationF2}
\end{equation}
which will be used in the rest of this appendix.

Equations \eqref{eq:MaxwellSolution} and \eqref{eq:EinsteinEquationF2}
are universal: they do not depend on the particular origin of the vacuum field $F$.
The $F$ field can be replaced by any conserved variable $q$,
as discussed in Ref.~\cite{KlinkhamerVolovik2008a}
(see also Ref.~\cite{KlinkhamerVolovik2016-brane}
for a $q$-field in an entirely
different context, namely that of freely suspended films).
Note that the role of the cosmological constant
in the Einstein gravitational
equation \eqref{eq:EinsteinEquationF2}
is played by the vacuum energy density
\beq
\label{eq:def-rhoV-F}
\rho_{V}(F) \equiv \epsilon(F)-\mu\, F\,.
\eeq
This confirms the general equation (\ref{eq:EquationOfStateQ}) for the class of considered quantum vacua with a conserved vacuum variable.
Recall that $\rho_{V}(F)$  belongs to the IR physics,
as distinct from $\epsilon(F)$ which is determined
by the UV degrees of freedom.
Remark also that, using the definition \eqref{eq:def-rhoV-F},
we can write the solved generalized Maxwell equation
\eqref{eq:MaxwellSolution} as
\beq
\label{eq:MaxwellSolution-with-rhoV}
 \frac{d\rho_{V}(F)}{d F}+\frac{R}{16\pi} \frac{dG^{-1}(F)}{d F} =0\,.
\eeq

Turning to cosmology, we use
the standard spatially flat Robertson--Walker
metric \eqref{eq:RW} with cosmic scale factor $a(t)$
and the matter energy-momentum tensor of a homogeneous
perfect fluid with energy density $\rho_{M}(t)$ and pressure $P_{M}(t)$.
The quantum-vacuum variable is also assumed to be
homogeneous, $F=F(t)$.
In addition, we define the usual Hubble parameter by
\begin{equation}
H(t)\equiv \frac{1}{a(t)}\,\frac{d a(t)}{d t}\,.
\label{eq:Hubble-def}
\end{equation}
From the Maxwell-type equation \eqref{eq:MaxwellSolution-with-rhoV},
we then have:
\beqa
&&
\frac{3}{8\pi} \frac{dG^{-1}(F)}{d F} \left(\frac{d H}{dt} +2\,H^{2} \right) =
 \frac{d \rho_{V}(F)}{d F} \,,
\label{eq:MaxwellFRW}
\eeqa
and from the Einstein-type
equation \eqref{eq:EinsteinEquationF2}:
\bsubeqs\label{eq:EinsteinFRW-all}
\beqa
&&
G^{-1}\,H^2 =
\frac{8\pi}{3}\,  \rho_\text{tot} - H \,\frac{d G^{-1}}{dt}\,,
\label{eq:EinsteinFRW-E00eq}
\\[2mm]
&&
G^{-1}\,\left( 2\,\frac{d H}{dt}+3\, H^2 \right)
=
-8\pi\,  P_\text{tot} - 2\,H \,\frac{d G^{-1}}{dt}
-\frac{d^2 G^{-1}}{dt^2}  \,,
\label{eq:EinsteinFRW-E11eq}
\eeqa
\esubeqs
with total energy density and total
pressure
\beq\label{eq:rho-total} \rho_\text{tot}\equiv
\rho_{V}+\rho_{M}\,,\quad P_\text{tot} \equiv
P_{V}+P_{M}\,,
\eeq
for the effective vacuum energy density
\beq
\rho_{V}(F)= -P_{V}(F) = \epsilon(F) -\mu F\,.
\label{eq:EinsteinFRW-rhoV}
\eeq

%%\newpage%%tmp
The cosmological equations \eqref{eq:MaxwellFRW} and \eqref{eq:EinsteinFRW-all} give immediately energy conservation
of the matter component,
\beq\label{eq:matter-energy-conservation}
\frac{d\rho_{M}}{d t} +3\, H\, \Big(P_{M}+\rho_{M}\Big) = 0 \,,
\eeq
as should be the case for a standard matter
field $\psi$~\cite{KlinkhamerVolovik2008b}.
Note, finally, that multiplication of the cosmological
Maxwell-type equation \eqref{eq:MaxwellFRW} by
$d F/d t$ gives
\begin{equation}
\frac{d \rho_{V}}{d t}
=
\frac{3}{8\pi}\, \frac{dG^{-1}}{d t}
\left( \frac{d H}{d t}
+2\,H^{2}\right)  \,.
\label{eq:MaxwellFRW-a-deriv}
\end{equation}
This last equation shows that
the cosmological ``constant'' (i.e., the vacuum energy density $\rho_{V}$) can change with cosmological time $t$, provided the
gravitational coupling depends on the
vacuum variable [$G=G(F)$ with $F=F(t)$].
For completeness, we remark that a variable
$\rho_{V}$ can also have other origins, such
as vacuum-matter energy exchange~\cite{KlinkhamerVolovik2016-CCP}
or higher-derivative terms~\cite{KlinkhamerVolovik2016-q-DM}.

%%\newpage%%tmp
\section{Analytic and numeric results}
\label{app:Analytic-and-numeric-results}

This appendix is more or less self-contained, but,
for a better understanding, it is advisable to first read the main text.

We start from the relatively simple theory
considered in Ref.~\cite{KlinkhamerVolovik2008b} and reviewed
in App.~\ref{subapp:Action-and-field-equations-of-q-theory}.
The action consists of three terms.
The first action term corresponds to a potential term $\epsilon(q)$
involving the 4-form field strength
[here, described by a (pseudo)scalar field $q(x)$].
The second term is the Einstein--Hilbert term with
a $q$-dependent gravitational coupling parameter $G(q)$.
The third term, finally, is the standard matter
action term without further dependence on $q$
(in principle, the parameters of the matter action
could have a $q$-dependence).

For our cosmological discussion,
we take the spatially flat Robertson--Walker
metric \eqref{eq:RW} with a cosmic scale factor $a(t)$
and a homogeneous-perfect-fluid energy-momentum tensor for
the matter component with energy density $\rho_{M}(t)$
and pressure $P_{M}(t)$.  The vacuum component is determined
by a homogeneous vacuum variable $q=q(t)$.

Dimensionless variables ($q_{0}=1$) are obtained as follows:
\bsubeqs\label{eq:dimensionless-variables}
%\beqa
\begin{align}
q &\to f\,,
\hspace*{-10mm}%\;\;
&G^{-1} &\to k\,,
\hspace*{-10mm}%\;\;
&H &\to h\,,
\\[2mm]
t &\to \tau\,,
\hspace*{-10mm}%\;\;
&\mu &\to u\,,
\hspace*{-10mm}%\;\;
&\rho_{V,\,M} &\to r_{V,\,M}\,.
\end{align}
%\eeqa
\esubeqs
Then, there are the following dimensionless  cosmological ordinary 
differential equations (ODEs)~\cite{KlinkhamerVolovik2008b}: 
\bsubeqs\label{eq:dimensionless-ODEs}
\beqa
\label{eq:dimensionless-ODEs-maxwell}
&&
r_{V}^{\prime} = k^{\prime}\,\big(\dot{h}+2\,h^2\big)\,,
\\[2mm]
\label{eq:dimensionless-ODEs-friedmann}
&&
r_{V}+r_M = k\,h^2+h\,\dot{k}\,,
\\[2mm]
\label{eq:dimensionless-ODE-M}
&&
\dot{r}_M + 3\,h\,(1+w_M)\,r_M = 0\,,
\\[2mm]
&&
r_{V} \equiv \epsilon - u\,f\,,
\eeqa
\esubeqs
where the prime denotes differentiation with respect to $f$
and the overdot differentiation with respect to $\tau$.
The above ODEs agree with those in Eqs.~(4.12abc)
of Ref.~\cite{KlinkhamerVolovik2008b}, as
we have absorbed a factor $3/(8\pi)$ into $k$.
[It is also obvious that the dimensionless ODEs
\eqref{eq:dimensionless-ODEs-maxwell},
\eqref{eq:dimensionless-ODEs-friedmann}, and
\eqref{eq:dimensionless-ODE-M}
correspond to
\eqref{eq:MaxwellFRW},
\eqref{eq:EinsteinFRW-E00eq}, and
\eqref{eq:matter-energy-conservation}
in App.~\ref{subapp:Action-and-field-equations-of-q-theory} above.]
In addition, we have used the dimensionless Hubble parameter,
\beq
\label{eq:hubble-def}
h(\tau) = \dot{a}(\tau)/a(\tau)\,,
\eeq
obtained from the cosmic scale factor $a(\tau)$ of the
spatially flat Robertson--Walker metric.

Next, make the following \textit{Ans\"{a}tze}:
\bsubeqs\label{eq:dimensionless-Ansaetze}
\beqa
\label{eq:dimensionless-Ansatz-k}
k(f) &=& \left(\left(f^2 \right)^{1/2}\right)^{1/2} = \sqrt{|f|}\,,
\\[2mm]
\label{eq:dimensionless-Ansatz-epsilon}
\epsilon(f) &=& -\frac{3}{2}\,f^2+\frac{1}{2}\,f^4\,,
\\[2mm]
\label{eq:dimensionless-Ansatz-u}
u(\tau) &=&
\begin{cases}
-1 \,,   &  \;\;\text{for}\;\;\tau \geq 0 \,,
\\[0mm]
\phantom{-}1 \,,   &  \;\;\text{for}\;\;\tau \leq 0  \,.
\end{cases}
\eeqa
\esubeqs
For the record, the \textit{Ans\"{a}tze}  
\eqref{eq:dimensionless-Ansatz-k},
\eqref{eq:dimensionless-Ansatz-epsilon} and
\eqref{eq:dimensionless-Ansatz-u}
correspond to
\eqref{eq:G},
\eqref{eq:epsilon}, and
\eqref{eq:NegT22} in the main text.
In \eqref{eq:dimensionless-Ansatz-u},
the double-valuedness of the chemical potential $u$
at $\tau=0$ (or $\mu$
at $t=0$ for the original quantities)
allows for a solution of the generalized Maxwell equation,
as given by \eqref{eq:Maxwell2}
in App.~\ref{subapp:Action-and-field-equations-of-q-theory}.
%%, with $q$ denoted by $F$ and a factor $8\pi/3$ absorbed into $G(q)$ here.
A further comment on the double-valuedness  
at $\tau=0$ will appear below.

\begin{figure}[t]
%\vspace*{-0mm}
\begin{center}
%%rename BB-as-TopQPT-2021-fig2-v2.eps  --> BB-as-TopQPT-fig2-v301.eps
\hspace*{0mm}
\includegraphics[width=1\textwidth]{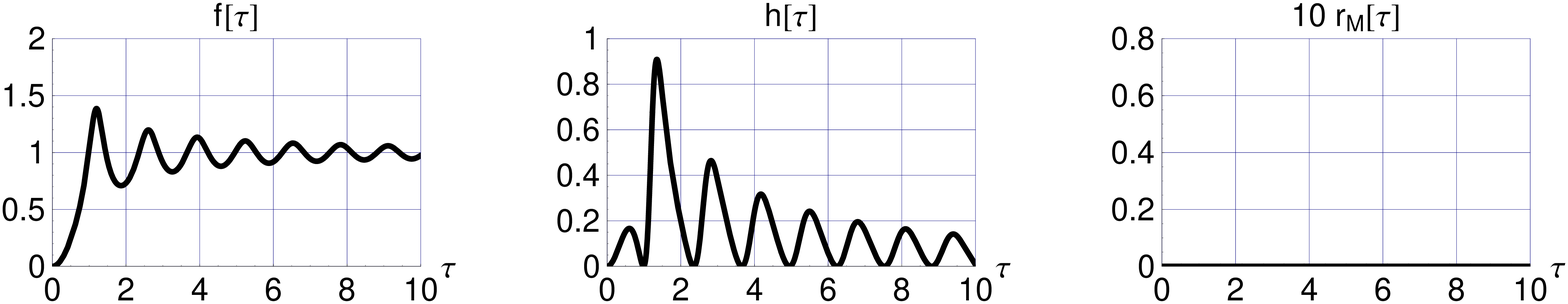}
%#{BB-as-TopQPT-fig2-v301.eps}
%%{BB-as-TopQPT-2021-fig2-v2.eps}
%%{BB-as-TopQPT-2021-fig01-v184.eps}
%%{BB-as-TopQPT-2021-fig2-v1.eps}
%%{BB-as-TopQPT-2021-fig01-v072.eps}
\end{center}\vspace*{-4mm}
\caption{
Positive-$\tau$ solution of the ODEs \eqref{eq:dimensionless-ODEs}
for $w_M=1/3$
with \textit{Ans\"{a}tze} \eqref{eq:dimensionless-Ansaetze} and
boundary conditions at $\tau_\text{bcs}=1/100$
from \eqref{eq:series-type-solution} with $\gamma=0$.
}
\label{fig:postbouncesol-nomatter}
\vspace*{0mm}
%\end{figure}
\vspace*{0mm}
%\begin{figure}[t]
\begin{center}
%%rename BB-as-TopQPT-2021-fig3-v2.eps  --> BB-as-TopQPT-fig3-v301.eps
\hspace*{0mm}
\includegraphics[width=1\textwidth]{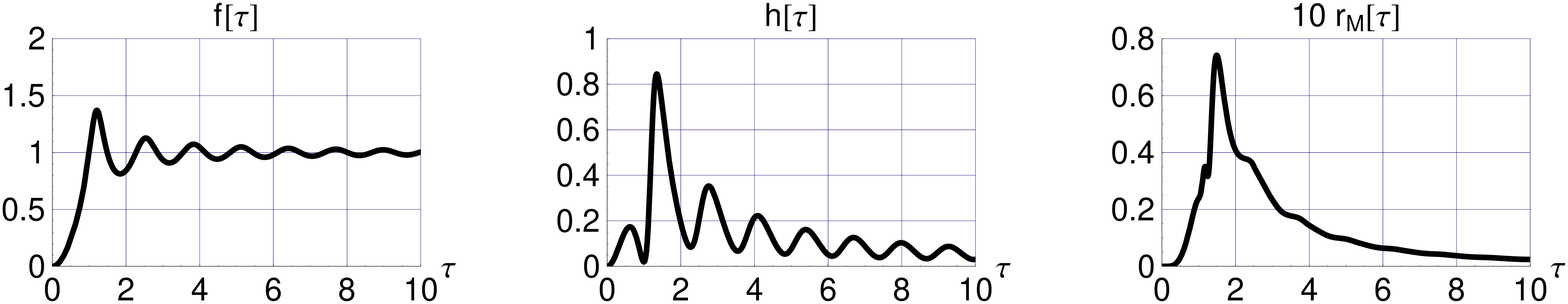}
%%{BB-as-TopQPT-fig3-v301.eps}
%%{BB-as-TopQPT-2021-fig3-v2.eps}
%%{BB-as-TopQPT-2021-fig02-v184.eps}
%%{BB-as-TopQPT-2021-fig3-v1.eps}
%%{BB-as-TopQPT-2021-fig02-v072.eps}
\end{center}\vspace*{-4mm}
\caption{Same as Fig.~\ref{fig:postbouncesol-nomatter}
but now from the ODEs \eqref{eq:dimensionless-ODEs-mod}
with parameters $\{w_M,\, \gamma \}= \{1/3,\, 1/5\}$.
}
\label{fig:postbouncesol-matter}
\vspace*{0mm}
%\end{figure}
\vspace*{0mm}
%\begin{figure}[t]
\begin{center}
%%rename BB-as-TopQPT-2021-fig4-v2.eps  --> BB-as-TopQPT-fig4-v301.eps
\hspace*{0mm}
\includegraphics[width=1\textwidth]{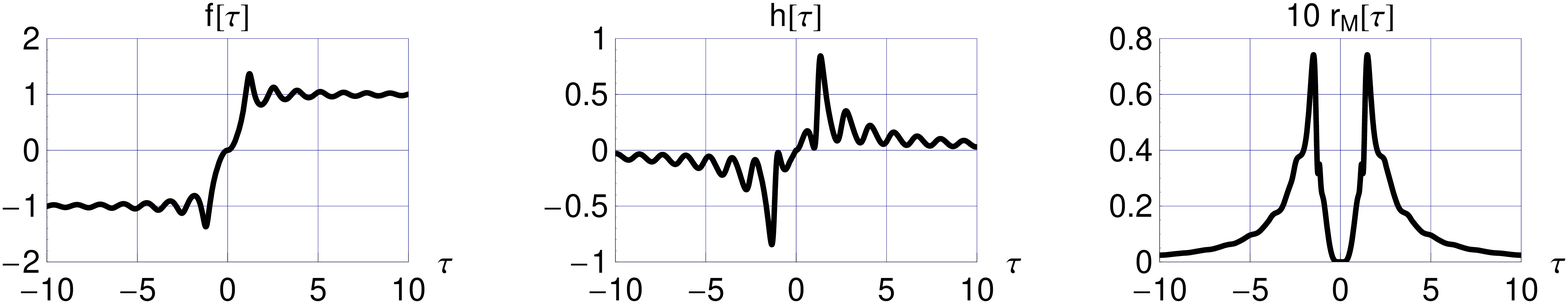}
%%{BB-as-TopQPT-fig4-v301.eps}
%%{BB-as-TopQPT-2021-fig4-v2.eps}
%%{BB-as-TopQPT-2021-fig03-v184.eps}
%%{BB-as-TopQPT-2021-fig4-v1.eps}
%%{BB-as-TopQPT-2021-fig03-v072.eps}
\end{center}\vspace*{-4mm}
\caption{Positive-$\tau$ numerical solution from
Fig.~\ref{fig:postbouncesol-matter}
and the corresponding negative-$\tau$ numerical solution from
boundary conditions at $\tau=-\tau_\text{bcs}=-1/100$.
The series-type solution  \eqref{eq:series-type-solution}  is plotted
for $\tau \in [-\tau_\text{bcs},\, \tau_\text{bcs}]$,
but this is barely visible for the $[-10,\,10]$ range shown
(see Fig.~\ref{fig:completesol-matter-central} for an enlargement).
}
\label{fig:completesol-matter}
\vspace*{0mm}
%\end{figure}
\vspace*{0mm}
%\begin{figure}[t]
\begin{center}
%%rename BB-as-TopQPT-2021-fig5-v2.eps  --> BB-as-TopQPT-fig5-v301.eps
\hspace*{0mm}
\includegraphics[width=1\textwidth]{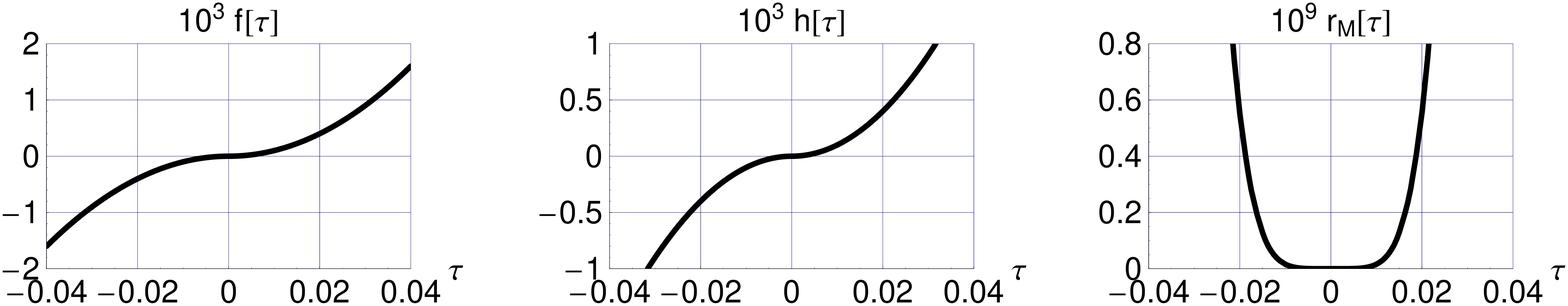}
%%{BB-as-TopQPT-fig5-v301.eps}
%%{BB-as-TopQPT-2021-fig5-v2.eps}
%%{BB-as-TopQPT-2021-fig04-v184.eps}
%%{BB-as-TopQPT-2021-fig5-v1.eps}
%%{BB-as-TopQPT-2021-fig04-v072.eps}
\end{center}\vspace*{-4mm}
\caption{Solution from Fig.~\ref{fig:completesol-matter}
plotted over the central region
$-4\,\tau_\text{bcs} \leq \tau \leq 4\,\tau_\text{bcs}$.
}
\label{fig:completesol-matter-central}
\vspace*{0mm}
\end{figure}

For $\tau \sim 0$, a series-type solution (denoted by a bar) reads
\bsubeqs\label{eq:series-type-solution}
\beqa
\label{eq:series-type-solution-f}
\overline{f}(\tau) &=& \tau\,|\tau|\,
\left(1+  \frac{1}{3}\,\tau^2 + \ldots\right)\,,
\\[2mm]
\label{eq:series-type-solution-h}
\overline{h}(\tau) &=& \tau\,|\tau|\,
\left(1-\frac{3}{2}\,\tau^2 - \frac{5}{8}\,\tau^4  +\ldots\right)\,,
\\[2mm]
\label{eq:series-type-solution-rM}
\overline{r}_M(\tau) &=& |\tau|\,
\left(0 + \frac{4}{5}\,\gamma\;\tau^4  + \ldots\right)\,,
\eeqa
\esubeqs
with coefficients obtained from the ODEs \eqref{eq:dimensionless-ODEs}
with \textit{Ans\"{a}tze}  \eqref{eq:dimensionless-Ansaetze}
and $\gamma=0$ for the moment (see below for further
discussion).
Numerical results for $\tau>0$ are shown in
Fig.~\ref{fig:postbouncesol-nomatter}.

These functions $\overline{f}$ and $\overline{h}$
have a discontinuous second derivative at $\tau=0$.
The second derivative of $\overline{f}$, in particular,
appears in the second Friedmann equation
[that equation is given
by \eqref{eq:EinsteinFRW-E11eq}
in App.~\ref{subapp:Action-and-field-equations-of-q-theory}].
However, it turns out that the dangerous term
$\ddot{f}$ appears in the combination
\mbox{$\big( -2\,\tau+|\tau|\,\ddot{f} \,\big)$},
so that the discontinuity is removed in this second Friedmann
equation for $f=\overline{f}  \sim \tau\,|\tau|$ at $\tau \sim 0$:
$\big( -2\,\tau+
|\tau|\,d^2 \overline{f}/d\tau^2 \,\big)\sim 0$.
Physically, this result is important for
energy conservation.

For completeness, we also give
the series-type solution (denoted by a bar) of
the cosmic scale factor $a(\tau)$.
From \eqref{eq:hubble-def},
\eqref{eq:series-type-solution-h},
and the boundary condition $a(0^{+})=1$,
there are then two types of solutions for $\overline{a}(\tau)$,
even and odd.
Specifically, we find for $\tau \sim 0$:
\bsubeqs\label{eq:abareven-abarodd}
\beqa
\label{eq:abareven}
\overline{a}_\text{even}(\tau) &=&
1
+ \frac{1}{3} \,|\tau|\,\tau^2
- \frac{3}{10}\,|\tau|\,\tau^4
+ \frac{1}{18}\,\tau^6
- \frac{5}{56}\,|\tau|\,\tau^6 + \ldots\,,
\\[2mm]
\label{eq:abarodd}
\overline{a}_\text{odd}(\tau) &=&
\overline{a}_\text{even}(\tau)
\,\times\,
\begin{cases}
\phantom{-}1 \,,   &  \;\;\text{for}\;\;\tau\geq 0 \,,
 \\[0mm]
-1 \,,   &  \;\;\text{for}\;\;\tau\leq 0  \,.
\end{cases}
\eeqa
\esubeqs
This last solution, as it stands, is double-valued  
at $\tau = 0$ and may possibly be relevant for the
universe-antiuniverse pair as discussed in
Sec.~\ref{subsec:Possible-physical-interpretation}.

According to \eqref{eq:series-type-solution-f}
and \eqref{eq:series-type-solution-h}, the
big bang is replaced by a topological quantum phase transition
in the vacuum and matter is perhaps generated by the oscillations
of the vacuum variable $f$ (cf. the left panel of
Fig.~\ref{fig:postbouncesol-nomatter}).
A possible term for vacuum-matter energy exchange gives
the following modified ODEs:
\bsubeqs\label{eq:dimensionless-ODEs-mod}
\beqa
\label{eq:dimensionless-ODEs-mod-rVdot}
&&
\dot{r}_{V} =
\dot{k}\,\big(\dot{h}+2\,h^2\big) - \gamma\,h\,\dot{f}^2\,,
\\[2mm]
&&
r_{V}+r_M = k\,h^2+h\,\dot{k}\,,
\\[2mm]
\label{eq:dimensionless-ODEs-mod-rMdot}
&&
\dot{r}_M + 3\,h\,(1+w_M)\,r_M = \gamma\,h\,\dot{f}^2\,,
\\[2mm]
&&
r_{V} \equiv \epsilon - u\,f\,,
\eeqa
\esubeqs
for $\gamma > 0$ and $\tau \ne 0$.
In order to arrive at \eqref{eq:dimensionless-ODEs-mod-rVdot},
we have multiplied \eqref{eq:dimensionless-ODEs-maxwell} by $\dot{f}$
and we only consider nonzero values of $\tau$,
for which $u(\tau)$ is constant.
Now, the series-type solution \eqref{eq:series-type-solution-rM}
picks up a nonzero quintic term in $\overline{r}_M$.
Numerical results for $\tau>0$ and $\gamma=1/5$
are shown in Fig.~\ref{fig:postbouncesol-matter}.

\begin{figure}[t]
\begin{center}
\vspace*{-2mm}
\hspace*{-0mm}
\includegraphics[width=1\textwidth]{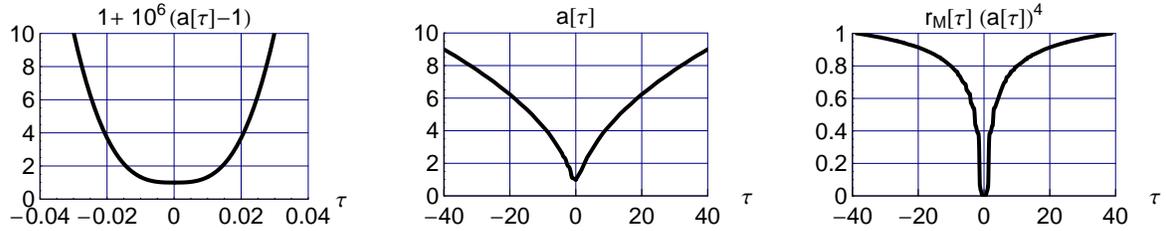}
%%{BB-as-TopQPT-fig6-v301.eps}
%%{BB-as-TopQPT-2021-fig6-v2.eps}
%%{BB-as-TopQPT-2021-fig05-v184.eps}
\end{center}
\vspace*{-6mm}
\caption{Even solution of the cosmic scale factor $a(\tau)$
corresponding to the solutions of
Fig.~\ref{fig:completesol-matter}, with
 the series-type solution \eqref{eq:abareven}
for $\tau \in [-\tau_\text{bcs},\, \tau_\text{bcs}]$
and matching boundary conditions on $a(\tau)$ at
$\tau=\pm \tau_\text{bcs}=\pm 1/100$.
The right panel shows the combination
$r_{M}(\tau) \, a^4(\tau)$.}
\label{fig:completesol-matter-a-rMa4}
\vspace*{-6mm}
\vspace*{0mm}
\end{figure}

This last numerical solution is plotted in
Fig.~\ref{fig:completesol-matter}
over the complete $\tau$
axis (or rather a finite segment thereof), with an enlargement  
of the central region in Fig.~\ref{fig:completesol-matter-central}.
The corresponding even solution of the
cosmic scale factor $a(\tau)$ is given
in Fig.~\ref{fig:completesol-matter-a-rMa4},
together with the combination $r_{M} \, a^4$,
which shows that the matter, after a creation phase,
dilutes more or less in the standard way
(now shown over a somewhat larger $\tau$ range).

From Fig.~\ref{fig:completesol-matter-a-rMa4}
for the even solution,
we see that $a(\tau) \geq 1$ for all values of $\tau$
and there is obviously
no geodetic incompleteness at $\tau=0$
for the spatially flat Robertson--Walker
metric \eqref{eq:RW}.
The same conclusion holds
for the odd solution \eqref{eq:abarodd},
as long as bosonic observables are considered.
Recall that geodetic incompleteness is the
defining characteristic of the Friedmann big bang
singularity~\cite{HawkingEllis1973}.

\end{appendix}

\newpage  %%tmp

\end{document}